\documentclass[11pt,superscriptaddress,aps,prd,preprint]{revtex4}
\usepackage[dvips]{graphicx}
\usepackage{amsfonts}
\newcommand{\be}{\begin{equation}}
\newcommand{\en}{\end{equation}}
\newcommand{\ba}{\begin{eqnarray}}
\newcommand{\ea}{\end{eqnarray}}

\newcommand{\Slash}[1]{{#1}\!\!\!/}
\newcommand{\RM}[1]{\mathrm{#1}}

\begin{document}

\title{Dynamical Lorentz and CPT symmetry breaking in a $4$D four-fermion model}

\author{M. Gomes}
\affiliation{Instituto de F\'{\i}sica, Universidade de S\~ao Paulo\\
Caixa Postal 66318, 05315-970, S\~ao Paulo, SP, Brazil}
\email{mgomes,ajsilva,tmariz,jroberto@fma.if.usp.br}
\author{T. Mariz}
\affiliation{Instituto de F\'{\i}sica, Universidade de S\~ao Paulo\\
Caixa Postal 66318, 05315-970, S\~ao Paulo, SP, Brazil}
\email{mgomes,ajsilva,tmariz,jroberto@fma.if.usp.br}
\author{J. R. Nascimento}
\affiliation{Instituto de F\'{\i}sica, Universidade de S\~ao Paulo\\
Caixa Postal 66318, 05315-970, S\~ao Paulo, SP, Brazil}
\affiliation{Departamento de F\'{\i}sica, Universidade Federal da Para\'{\i}ba\\
Caixa Postal 5008, 58051-970, Jo\~ao Pessoa, Para\'{\i}ba, Brazil}
\email{mgomes,ajsilva,tmariz,jroberto@fma.if.usp.br}
\author{A. J. da Silva}
\affiliation{Instituto de F\'{\i}sica, Universidade de S\~ao Paulo\\
Caixa Postal 66318, 05315-970, S\~ao Paulo, SP, Brazil}
\email{mgomes,ajsilva,tmariz,jroberto@fma.if.usp.br}

\begin{abstract}
In a $4$D chiral Thirring model we analyse the possibility that radiative corrections may produce spontaneous breaking of Lorentz and CPT symmetry. 
By studying the effective potential, we verified that the chiral current $\bar\psi\gamma^{\mu} \gamma_5 \psi$ may assume a nonzero vacuum expectation value which triggers Lorentz and CPT violations. Furthermore, by making fluctuations on the minimum of the potential we  dynamically induce
a bumblebee like model  containing a Chern-Simons term.
\end{abstract}


\maketitle

\section{Introduction}

The Lorentz invariance is one of the most well established symmetries in physics having survived a variety of stringent tests.
 Nevertheless, recently there has been an active interest on the possibility that more fundamental theories may induce small violations of Lorentz invariance into the standard model, at levels accessibles to high precision experiments \cite{Kos}. The original motivation for this idea arose from the fact that the spontaneous breaking of Lorentz symmetry may appear in the context of string theory \cite{KosSam} (in field theory the breaking was first studied in \cite{CarFie}). To
systematically investigate this possibility, a standard model extension (SME) including all possible terms which may violate Lorentz and/or CPT invariance, was constructed \cite{SME}. 

The breaking of the Lorentz symmetry in the SME was generated by a procedure analogous to the Higgs mechanism in which a scalar field gains a vacuum expectation value (VEV) to furnish masses for the standard model particles. Nonzero expectation values for tensor fields that contain Lorentz indices select specific directions in the spacetime, breaking Lorentz invariance spontaneously. As an example, let us consider a toy model whose Lagrangian describes a vector field $B_\mu$ in such way to induce spontaneous Lorentz and CPT violation \cite{KosLeh,Alt,Ber}, 
\be\label{toy}
{\cal L} = -\frac14 F_{\mu\nu}F^{\mu\nu} + \bar\psi(i\Slash{\partial} - m\,-\, e\Slash{B}\gamma_5)\psi - \frac14\lambda(B_\mu B^\mu - \beta^2)^2,
\en
where $F_{\mu\nu}=\partial_\mu B_\nu - \partial_\nu B_\mu$.  The Maxwell form of the kinetic part of $B_\mu$ can be justified by energy considerations \cite{Chk} without recourse to a gauge invariance principle. The self-interaction in this ``bumblebee'' model triggers a Lorentz and CPT-violating VEV $\left\langle B_\mu \right\rangle = \beta_\mu$. Very interesting terms are obtained when we consider fluctuations about the vacuum through the redefinition $B_\mu = \beta_\mu + A_\mu$, where the shifted field is assumed to have a zero VEV, $\left\langle A_\mu \right\rangle = 0$.  The Lagrangian (\ref{toy}) becomes
\be
{\cal L} = -\frac14 F_{\mu\nu}F^{\mu\nu} + \bar\psi(i\Slash{\partial} - m\,-\, \Slash{b}\gamma_5 - e\Slash{A}\gamma_5)\psi - \frac14\lambda\left(A_\mu A^\mu - \frac2e A\cdot b\right)^2,
\en
with $b_\mu = e\beta_\mu$, presenting the term $b_\mu\bar\psi\gamma^\mu\gamma_5\psi$ which violates the Lorentz and CPT symmetry. This term can be used to produce through radiative corrections the Chern-Simons Lagrangian \cite{1},  
\be
{\cal L}_\mathrm{CS} = \textstyle\frac12\kappa^\mu\epsilon_{\mu\nu\lambda\rho}A^\nu F^{\lambda\rho},
\en
with $\kappa_\mu \propto b_\mu$, since they have the same C, P and T transformation properties. Both at zero \cite{1,2,3,4,5,6,7,8,9,10,11,12,13,14,15} and at finite temperature \cite{Nas,Cer,Ebe,Mar,Nas2},  in the non-Abelian case \cite{Gom}, and in contexts which include gravity \cite{Mar2,Mar3}, this issue has been carefully investigated.

In the present work, we will analyze the spontaneous breaking of Lorentz and CPT symmetry \cite{And} via the Coleman-Weinberg mechanism \cite{ColWei}. Our objective is to examine the possibility of causing a spontaneous Lorentz and CPT symmetry breaking through radiative corrections starting from the self-interacting fermionic theory given by the Lagrangian
\begin{equation}\label{tThirring}
{\cal L}_0 = \bar\psi(i\Slash{\partial}-m)\psi - \frac G2 (\bar\psi\gamma_\mu\gamma_5\psi)(\bar\psi\gamma^\mu\gamma_5\psi),
\end{equation}
and   dynamically inducing a  bumblebee model with a Chern-Simons  term. A similar mechanism was proposed long time ago \cite{Bjo} as a way to generate the quantum electrodynamics (QED) through radiative corrections without invoking local $U(1)$ gauge invariance \cite{Bia,Gur,Egu}. For some recent developments see \cite{Bjo2,Kra,Jen}. 

The model given by (\ref{tThirring}) is non-renormalizable and must be thought as a low energy effective theory arising from a more fundamental, yet unknown theory, in the same sense as the original proposal of Nambu and Jona-Lasinio (NJL) \cite{Nam} for QCD. As in the NJL model an UV cutoff will be present in the results, which will represent our lack of knowledge of the physics beyond that scale. In fact, we will use a variant of the dimensional regularization prescription and the parameter $\epsilon=4-D$ will be present (a correspondence between $\epsilon$ and a momentum cutoff $\Lambda$ is discussed in many places in the literature \cite{Aka,Aka2}).

This paper is organized as follows. In the Section \ref{EffPot} we  show that  a Higgs-like potential may be induced through radiative corrections from the Lagrangian (\ref{tThirring}), instead of been added from the start as in the bumblebee model (\ref{toy}), leading to the appearance of a Lorentz- and CPT-violating VEV $\langle \bar\psi\gamma_\mu\gamma_5\psi \rangle \neq 0$. After taking into account fluctuations about this vacuum, the radiative corrections at one-loop are examined in Section \ref{OneLoop}. Section \ref{Conc} contains some final comments. 

\section{Effective potential}\label{EffPot}

\begin{figure}[h]
	\centering
		\includegraphics{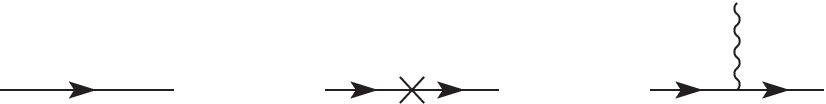}
  \caption{Feynman rules. Continuous and wave lines represent the fermion propagator and the auxiliary field, respectively. The cross indicates the $-i\Slash{b}\gamma_5$ insertion in the fermion propagator and the trilinear vertex corresponds to $-i e \gamma^\mu\gamma^5$}
	\label{fig1}
\end{figure}

\begin{figure}[h]
	\centering
		\includegraphics{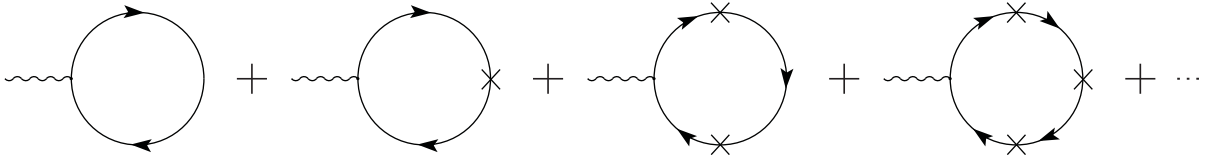}
	\caption{Contributions to the tadpole $\Pi^\mu$}
	\label{fig2}
\end{figure}

In order to eliminate the self-interaction term of Eq.~(\ref{tThirring}), it is convenient to introduce an auxiliary field $B_{\mu}$, so that the above Lagrangian can be rewritten as
\begin{eqnarray}\label{Lag}
{\cal L} &=& {\cal L}_0 + \frac{g^2}{2}\left(B_\mu - \frac{e}{g^2} \bar\psi\gamma_\mu\gamma_5\psi\right)^2 \nonumber\\
 &=& \frac{g^2}{2}B_\mu B^\mu + \bar\psi(i\Slash{\partial} - m\,-\, e\Slash{B}\gamma_5)\psi
\end{eqnarray}
where $G=e^2/g^2$. To verify the possibility that a bumblebee potential can be induced through radiative corrections from this Lagrangian, we consider the generating functional defined as
\begin{equation}
Z(\bar \eta,\,\eta) = \int DB_\mu D\psi D\bar\psi e^{i\int d^4x({\cal L}+\bar\eta\psi+\bar\psi\eta)}.
\end{equation}
By performing the fermionic integration we get
\be\label{GF}
Z(\bar \eta,\,\eta) = \int DB_\mu \exp\left[iS_\RM{eff}[B] + i\int d^4x \left(\bar\eta\frac{1}{i\Slash{\partial}-m\,-\,e\Slash{B}\gamma_5}\eta \right) \right],
\en
where the effective action is given by
\begin{equation}\label{1}
S_\RM{eff}[B] = \frac{g^2}{2} \int d^4x\, B_\mu B^\mu -i \RM{Tr} \ln(i\Slash{\partial}-m\,-\,e\Slash{B}\gamma_5). 
\end{equation}
The $\RM{Tr}$ stands for the trace over Dirac matrices as well as the trace over the integration in momentum or coordinate spaces. Thus, the effective potential turns out to be
\be\label{Vef}
V_\RM{eff} = -\frac{g^2}{2}B_\mu B^\mu + i\,\RM{tr} \int\frac{d^4p}{(2\pi)^4}\, \ln(\,\Slash{p}-m\,-\,e\Slash{B}\gamma_5),
\en
where the classical field is in a coordinate independent configuration. As we are interested in verifying the existence of a nontrivial minimum, we look for solutions of the expression
\be\label{DVef}
\frac{dV_\RM{eff}}{dB_\mu}\Big|_{B=\beta} =  - \frac{g^2}e b^\mu - i\,\Pi^\mu = 0,
\en
where $b^\mu=e\beta^\mu\not= 0$ and $\Pi^\mu$ is the one-loop tadpole amplitude: 
\begin{equation}\label{Pi}
\Pi^\mu = \RM{tr} \int\frac{d^4p}{(2\pi)^4} \frac{i}{\,\Slash{p}-m\,-\,\Slash{b}\gamma_5}(-ie) \gamma^\mu\gamma_5.
\end{equation}
To evaluate this integral we will follow the perturbative route where now the propagator is the usual $S(p)=i(\Slash{p}-m)^{-1}$ and $-i\Slash{b}\gamma_5$ is considered as insertions in this propagator. At this point a graphical representation may be helpful. With the conventions indicated in Fig. \ref{fig1} the contributions to $\Pi^{\mu}$ are shown in Fig \ref{fig2}. Our regularization procedure, the dimensional reduction scheme \cite{Buras}, consist in calculating the traces of the Dirac matrices in 4 dimensions and afterwards promoting the metric tensor $g^{\mu\nu}$ and the integrals to D dimensions. Proceeding in this way, we found that the first and third graphs as well as graphs with more than three insertions vanish \footnote{By following a technique similar to the one in Ref. \cite{3}, using cutoff regularization we verified that higher powers than three in $b_\mu$ vanish.}. The remaining contributions, i.e.,
the second and fourth graphs, give
\begin{equation}\label{Pimu}
\Pi^\mu = \left[-\frac{im^2e}{\pi^2\epsilon}+\frac{im^2e}{2\pi^2}\ln\left(\frac{m^2}{\mu'^2}\right)-\frac{ib^2e}{3\pi^2}\right]b^\mu,
\end{equation}
with $\epsilon = 4-D$, $\mu'^2=4\pi\mu^2e^{-\gamma}$, and $\mu$ been the renormalization spot. Then, the expression (\ref{DVef}) can be rewritten as 
\be\label{DVef2}
\left[-\frac1G_\mathrm{R}+\frac{m^2}{2\pi^2}\ln\left(\frac{m^2}{\mu'^2}\right)-\frac{b^2}{3\pi^2}\right]eb_\mu = 0,
\en
where we have introduced the renormalized coupling constant
\be
\frac1G_\mathrm{R} = \frac1G + \frac{m^2}{\pi^2\epsilon}.
\en
Therefore, we see that a nontrivial solution of this gap equation is
\be\label{min}
b^2 = -3\pi^2\left[\frac1{G_\mathrm{R}} - \frac{m^2}{2\pi^2}\ln\left(\frac{m^2}{\mu'^2}\right)\right].
\en
 From this equation we see that a nontrivial minimum  with a timelike $b_\mu$ is possible if
\be\label{cond1}
G_\RM{R} > \frac{2\pi^2}{m^2\ln\left(\frac{m^2}{\mu'^2}\right)},
\en
whereas  a nonzero spacelike $b_\mu$  requires
\be\label{cond2}
G_\RM{R} < \frac{2\pi^2}{m^2\ln\left(\frac{m^2}{\mu'^2}\right)}.
\en
The situation we are interested is the case where the effective potential  possess a nonzero minimum given by equation (\ref{min}), and therefore a VEV breaks the Lorentz invariance, i.e., $\langle B_\mu \rangle = \beta_\mu\not =0$.  This breaking of Lorentz invariance implies in a modification of the dispersion relation which may be useful in the study of ultra-high energy cosmic rays \cite{ColGla,ColGla2}.

\section{One-loop corrections and The induced Chern-Simons term}\label{OneLoop}

Let us now study the fluctuations, $B_\mu = \beta_\mu + A_\mu$,  around the nontrivial minimum of the potential. We anticipate that, due to the breaking of the Lorentz and CPT symmetry, Chern-Simons terms will occur. The generating functional (\ref{GF}) expressed in terms of the shifted field is
\be
Z(\bar \eta,\,\eta) = \int DA_\mu \exp\left[iS_\RM{eff}[A,b] + i\int d^4x \left(\bar\eta\frac{1}{i\Slash{\partial}-m\,-\,\Slash{b}\gamma_5-e\Slash{A}\gamma_5}\eta \right) \right],
\en
 where the effective action is given by
\begin{equation}
S_\RM{eff}[A,b] = \int d^4x\left(\frac{g^2}{2}A_\mu A^\mu+\frac{g^2}{e}A_\mu b^\mu+\frac{g^2}{2e^2}b_\mu b^\mu\right)-i \RM{Tr} \ln(i\Slash{\partial}-m\,-\,\Slash{b}\gamma_5-e\Slash{A}\gamma_5).
\end{equation}
Up to a field independent factor which may be absorbed in the normalization of the generating functional, we get
\begin{equation}
S'_\RM{eff}[A,b] = \int d^4x\left(\frac{g^2}{2}A_\mu A^\mu+\frac{g^2}{e}A_\mu b^\mu\right) + S^{(n)}_\RM{eff}[A,b],
\end{equation}
where
\begin{equation}\label{series}
S^{(n)}_\RM{eff}[A,b] = i \RM{Tr} \sum_{n=1}^\infty\frac1n\left[\frac i{i\Slash{\partial}-m\,-\,\Slash{b}\gamma_5}(-ie)\Slash{A}\gamma_5\right]^n.
\end{equation}
The formally divergent contributions in this formula are the tadpole, the self-energy,  the three and four point vertex functions of the  field $A_\mu $.  The tadpole is given by 
\begin{eqnarray}
S_\RM{eff}^{(1)}[A,b] &=& i \RM{Tr} \frac i{i\Slash{\partial}-m\,-\,\Slash{b}\gamma_5}(-ie)\Slash{A}\gamma_5 \nonumber\\
                      &=& i\int d^4x\, \Pi^\mu A_\mu,
\end{eqnarray}
where $\Pi^\mu$ was given in (\ref{Pimu}) due to (\ref{DVef}).

\begin{figure}[h]
	\centering
		\includegraphics{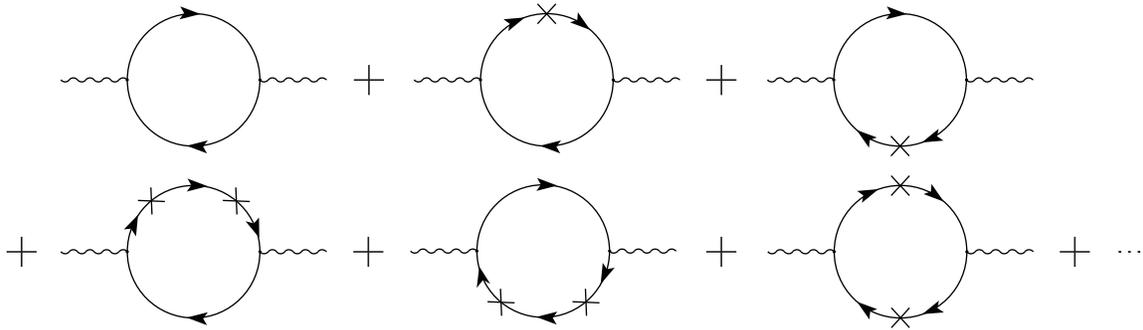}
  \caption{Contributions to the vacuum polarization $\Pi^{\mu\nu}$}
	\label{fig3}
\end{figure}

The self-energy term, which corresponds to $n=2$, yields
\begin{eqnarray}
S_\RM{eff}^{(2)}[A,b] &=& \frac i2 \RM{Tr} \frac i{i\Slash{\partial}-m\,-\,\Slash{b}\gamma_5}(-ie)\Slash{A}\gamma_5 \frac i{i\Slash{\partial}-m\,-\,\Slash{b}\gamma_5}(-ie)\Slash{A}\gamma_5 \nonumber\\
                      &=& \frac i2 \int d^4x\, \Pi^{\mu\nu} A_\mu A_\nu,
\end{eqnarray}
where
\begin{equation}\label{Pimunu}
\Pi^{\mu\nu} = \RM{tr} \int \frac{d^4p}{(2\pi)^4} \frac{i}{\,\Slash{p}-m\,-\,\Slash{b}\gamma_5}(-ie)\gamma^\mu\gamma_5 \frac{i}{\,\Slash{p}\,-\,i\Slash{\partial}-m\,-\,\Slash{b}\gamma_5}(-ie)\gamma^\nu\gamma_5.
\end{equation}
By expanding in powers of $\,\Slash{b}\gamma_5$, the above result can be expressed graphically as in Fig. \ref{fig3}. The second and third graphs are separately finite and furnish a nonlocal Chern-Simons term. Similarly to what happens in extended QED \cite{Jac,Per,Che} the coefficient of this generated Chern-Simons term is ambiguous, i.e., different regularizations produce distinct results; for example, by using the 't Hooft-Veltman prescription \cite{tHo,Bre} the coefficient vanishes. The divergent parts of the fourth, fifth, and sixth graphs cancel among themselves  (we have also verified that graphs with three and four insertions of the vertex $-i\Slash{b}\gamma_5$ vanish); so  only the first graph turns out to be divergent. We get 
\begin{eqnarray}
\Pi^{\mu\nu} &=& ie^2 g^{\mu\nu}\left[-\frac{m^2}{\pi^2\epsilon}+\frac{m^2}{2\pi^2}\ln\left(\frac{m^2}{\mu'^2}\right)-\frac{b^2}{3\pi^2}\right] - \frac{ie^2}{6\pi^2\epsilon} (g^{\mu\nu}\Box-\partial^\mu \partial^\nu) \\ 
 &+& \frac{ie^2}{12\pi^2}\left[\ln\left(\frac{m^2}{\mu'^2}\right)+1\right] (g^{\mu\nu}\Box-\partial^\mu \partial^\nu) - \frac{ie^2}{6\pi^2} \epsilon^{\mu\nu\lambda\rho}b_\lambda \partial_\rho - \frac{ie^2}{12\pi^2}\partial^\mu \partial^\nu  - \frac{2ie^2}{3\pi^2} b^\mu b^\nu, \nonumber
\end{eqnarray}
valid for $\Box/m^2<<1$.  

\begin{figure}[h]
	\centering
		\includegraphics{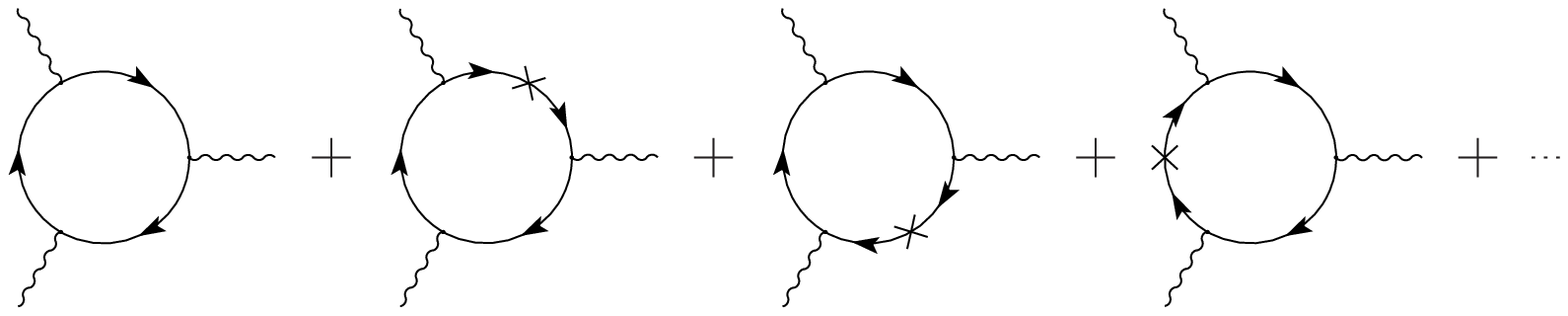}
	\caption{Contributions to the three-point $\Pi^{\mu\nu\rho}$}
	\label{fig4}
\end{figure}

Notice that ultraviolet (UV) divergences may also appear in the third term of the series in Eq. (\ref{series}), as Furry theorem is not applicable. For $n = 3$ the expression (\ref{series})
gives
\begin{eqnarray}
S_\RM{eff}^{(3)}[A,b] &=& \frac i3 \RM{Tr} \frac i{i\Slash{\partial}-m\,-\,\Slash{b}\gamma_5}(-ie)\Slash{A}\gamma_5 \frac i{i\Slash{\partial}-m\,-\,\Slash{b}\gamma_5}(-ie)\Slash{A}\gamma_5 \frac i{i\Slash{\partial}-m\,-\,\Slash{b}\gamma_5}(-ie)\Slash{A}\gamma_5 \nonumber\\
&=& \frac i3 \int d^4x\, \Pi^{\mu\nu\rho} A_\mu A_\nu A_\rho,
\end{eqnarray}
where
\begin{eqnarray}\label{Pimnr}
\Pi^{\mu\nu\rho} &=& \RM{tr} \int \frac{d^4p}{(2\pi)^4} \frac{i}{\,\Slash{p}-m\,-\,\Slash{b}\gamma_5}(-ie) \gamma^\mu\gamma_5 \frac{i}{\,\Slash{p}\,-\,i\Slash{\partial}-m\,-\,\Slash{b}\gamma_5}(-ie) \gamma^\nu\gamma_5 \nonumber\\ && \times \frac{i}{\,\Slash{p}\,-\,i\Slash{\partial}\,-\,i\Slash{\partial}'-m\,-\,\Slash{b}\gamma_5}(-ie) \gamma^\rho\gamma_5,
\end{eqnarray}
which, as a power series in $\Slash{b}\gamma_5$, is given by the graph expansion of Fig. \ref{fig4}. In the above formula  the derivatives $\Slash{\partial}$ and $\Slash{\partial}'$ act on $A_\mu$ and $A_\nu$, respectively. Due to properties of the trace of Dirac matrices the first graph results finite, whereas the divergent parts of the second, third, and fourth graphs cancel among themselves,  in the same way as what happens with some one-loop contributions to Lorentz-violating QED \cite{KosLan}.  The leading terms in the expansion in $\Box/m^2$  yields
\begin{equation}
\Pi^{\mu\nu\rho} = \frac{ie^3}{12\pi^2}(\epsilon^{\mu\nu\rho\lambda}\partial_\lambda - \epsilon^{\mu\nu\rho\lambda}\partial'_\lambda) +\frac{ie^3}{3\pi^2}(g^{\mu\nu}b^\rho + g^{\mu\rho}b^\nu + g^{\nu\rho}b^\mu).
\end{equation}

In principle the fourth term of the series in (\ref{series}) may be divergent but it results finite since the leading term is similar to the one in QED where as it is known, it is finite. We obtain
\be
S^{(4)}_\RM{eff} = \frac{e^4}{12\pi^2}\int d^4x\,(A_\mu A^\mu)^2 + {\cal O}\left(\frac\Box{m^2}\right).
\en

The results obtained so far allow us to write the effective Lagrangian as
\begin{eqnarray}\label{lag}
{\cal L} &=& -\frac1{4Z_3}F_{\mu\nu}F^{\mu\nu} + \frac{e^2}{24\pi^2}b^\mu\epsilon_{\mu\nu\lambda\rho}A^\nu F^{\lambda\rho} - \frac{e^2}{24\pi^2}(\partial_\mu A^\mu)^2 + \frac{e^4}{12\pi^2}\left(A_\mu A^\mu - \frac2e A\cdot b\right)^2 \nonumber\\
&&+\frac{e}{2b^2}A_\mu A^\mu \left\langle A_\nu \right\rangle b^\nu + \left\langle A_\mu \right\rangle A^\mu,
\end{eqnarray}
where
\begin{equation}\label{Z3}
\frac1{Z_3} = \frac{e^2}{6\pi^2\epsilon} - \frac{e^2}{12\pi^2}\left[\ln\left(\frac{m^2}{\mu'^2}\right)+1\right],
\end{equation}
and
\begin{eqnarray}
\left\langle A_\mu \right\rangle = \left[\frac1G_\mathrm{R}-\frac{m^2}{2\pi^2}\ln\left(\frac{m^2}{\mu'^2}\right)+\frac{b^2}{3\pi^2}\right]eb_\mu.
\end{eqnarray}
The requirement that  $\left\langle A_\mu \right\rangle = 0$, such that $B_\mu$ acquires a VEV $\left\langle B_\mu \right\rangle \neq 0$, was already studied in Eqs.~(\ref{DVef})-(\ref{min}), with the solutions (\ref{cond1}) and (\ref{cond2}). By defining a renormalized field $A_R^\mu = Z_3^{-1/2}A^\mu$ and a renormalized coupling constant $e_R = Z_3^{1/2} e$, we get 
\be\label{L}
{\cal L} = -\frac14 F_{\RM{R}\mu\nu}F_\RM{R}^{\mu\nu} + \frac{e_\RM{R}^2}{24\pi^2}b^\mu\epsilon_{\mu\nu\lambda\rho}A_\RM{R}^\nu F_\RM{R}^{\lambda\rho} - \frac{e_\RM{R}^2}{24\pi^2}(\partial_\mu A_\RM{R}^\mu)^2 + \frac{e_\RM{R}^4}{12\pi^2}\left(A_{\RM{R}\mu} A_\RM{R}^\mu - \frac2{e_\RM{R}} A_\RM{R}\cdot b\right)^2.
\en
This Lagrangian is exactly the extended QED by the Chern-Simons term, added of a gauge-fixing term and of a potential that do not trigger a Lorentz and CPT-violation. We should stress that the (finite) Chern-Simons coefficient is ambiguous and depends on the particular regularization scheme used \cite{Jac,Per,Che}.

By substituting the expression (\ref{Z3}) ($Z_3 \cong 6\pi^2\epsilon/e^2$) into the renormalized coupling constant, we obtain the result $e_\RM{R}^2 \cong 6\pi^2\epsilon$ which is the same one for the induced QED \cite{Bjo,Bia,Bjo2,Aka,Aka2}.  In the limit $\epsilon\to0$ we  would have a trivial free theory with vanishing coupling constant.  But as we remarked in the introduction we  must
keep $\epsilon$ at some small but nonvanishing value so that
 Eq. (\ref{L}) has to be interpreted as an effective theory. Bumblebee models of this type  have been discussed in flat and curved spacetime \cite{Kos2,BluKos}.

\section{Conclusions}\label{Conc}

 We have shown that a bumblebee potential can be induced through radiative corrections from a $4$D chiral Thirring model, as the conditions (\ref{cond1}) and (\ref{cond2})  hold for timelike and spacelike $b_\mu$, respectively. By considering the fluctuations on the minimum of the potential, the QED extended by the Chern-Simons term is dynamically generated.

{\bf Acknowledgements.} Authors are grateful to Prof.~V.~Alan~Kosteleck\'y for some enlightenments. This work was partially supported by Funda\c{c}\~{a}o de Amparo \`{a} Pesquisa do Estado de S\~{a}o Paulo (FAPESP) and Conselho Nacional de Desenvolvimento Cient\'{\i}fico e Tecnol\'{o}gico (CNPq). The work by T.~M. has been supported by FAPESP, project 06/06531-4.

\newpage

\end{document}